\documentclass[preprint]{elsarticle} 

\usepackage{lineno,hyperref}
\usepackage{amssymb, natbib}
\setcitestyle{authoryear,open={(},close={)}} 

\usepackage{graphics}
\usepackage{booktabs,caption}
\usepackage{threeparttable}
\newcommand{\msunyr}{M$_{\odot}$\,yr$^{-1}$}
\def\astrobj#1{#1}

\journal{New Astronomy}

\begin{document}

\begin{frontmatter}

\title{Automatic Detection of Accretion Bursts in Young Stellar Objects: a New Algorithm for Long--Term Sky Surveys} 





\author[address1,address2]{F. Strafella\corref{mycorrespondingauthor}}
\cortext[mycorrespondingauthor]{Corresponding author}
\ead{francesco.strafella@le.infn.it}

\author[address2,address3]{G. Altavilla}
\ead{giuseppe.altavilla@inaf.it}

\author[address2]{T. Giannini}
\ead{teresa.giannini@inaf.it}

\author[address3]{A. Giunta}
\ead{alessio.giunta@gmail.com}

\author[address2]{D. Lorenzetti}
\ead{dario.lorenzetti@inaf.it}

\author[address1,address2]{A. Nucita}
\ead{achille.nucita@le.infn.it}
\author[address1,address2]{A. Franco}
\ead{antonio.franco@le.infn.it}

\address[address1]{Dipartimento di Matematica e Fisica, Universit\`a del Salento \\
                    and INAF-Sezione di Lecce \\
                    via per Arnesano km.5, I-73100 Lecce, Italy}
\address[address2]{INAF - Osservatorio Astronomico di Roma, Via Frascati 33 \\ I-00078 Monte Porzio Catone (Roma), Italy}
\address[address3]{Space Science Data Center - ASI \\ Via del Politecnico snc, I-00133 Roma, Italy}

\begin{abstract}
Young stellar objects in their pre-main sequence phase are characterized by 
irregular changes in brightness, generally attributed to an increase of the mass 
accretion rate due to various kind of instabilities occurring in the circumstellar 
disk. In the era of large surveys aimed to monitor the sky, we present a 
pipeline to detect irregular bursts, in particular EXors-like ( EX Lupi type eruptive variables), 
in the light curves.
The procedure follows a heuristic approach and is tested against the light curves 
already collected for a few objects presently recognized as {\em bona fide} or candidate EXors. 
\end{abstract}

\begin{keyword}
stars: pre-main-sequence;  stars: variables: other;  methods: data analysis
\MSC[2010] 00-01\sep  99-00
\end{keyword}

\end{frontmatter}


\section{Introduction} \label{sec:intro}
Young stellar objects (YSOs) of low mass (0.5--2 M$_\odot$) accumulate about 90\% of their 
final mass in about 10$^5$ yr. After this stage, the mass accretion process continues for 
about 10$^7$ yr at lower accretion rates (around 10$^{-8}-10^{-7}$ \msunyr) as typical 
T Tauri stars. According to a widely accepted picture, matter in accretion viscously 
migrates through the circumstellar disk toward the inner edge from where it is intermittently 
channelled along the magnetic field lines to the central star; the fall onto the stellar 
surface produces a shock that cools by emitting a hot continuum \citep{Shu94}. 

Observationally, this kind of accretion manifests itself through sudden and intermittent bursts 
of luminosity as a consequence of the shock cooling, and produces a small and irregular 
photometric variability that typically ranges between 0.1$-$1 mag.

However, some young sources display more powerful outbursts with amplitudes of several visual magnitudes. 
These objects, generally known as eruptive variables, are usually classified into two 
main classes \citep[see also the review by][]{Aud14}:\\
(1) FUors \citep{Har85} characterized by bursts ($\Delta m_{\textrm{V}} \gtrsim 4$ mag) of long duration (tens of years) 
with accretion rates of the order of 10$^{-5}-10^{-4}$ M$_{\odot}$~yr$^{-1}$ and spectra constituted  by 
absorption lines;\\ 
(2) EXors \citep{Her89} with shorter outbursts ($2\lesssim \Delta m_{\textrm{V}}\lesssim 4$ mag, duration of 
months--one year) with a recurrence time of years, showing accretion rates of the order of 10$^{-7}-10^{-6}$~M$_{\odot}$~yr$^{-1}$, 
and characterized by emission line spectra \citep[e.g.][]{Her08,Lor09,Kos11,Sic12,Ant13}.\\
Unfortunately, only a handful of objects are considered as genuine members of each class \citep{Aud14}, although 
several efforts have been done so far either scrutinizing different catalogs \citep{Gia09, 
Sch13, Ant14}, or undertaking specific long-term monitoring programs \citep{Gra07, Gra08, Mor11, Hil13, Gia17}.

In the last decade, however, the situation has rapidly changed because multi-wavelength sky surveys 
have allowed to significantly increase the number of known or candidate eruptive variables. Among the ground-based surveys, 
which typically observe the optical and near-infrared sky with cadence of days or weeks, 
we underline the {\it All-Sky Automated Survey for Supernovae} (ASAS-SN)\footnote{http://www.astronomy.ohio-state.edu/asassn/index.shtml}, 
the {\it Intermediate Palomar Transient Factory} (iPTF)\footnote{https://www.ptf.caltech.edu/iptf}, 
the {\it Zwicky Transient Facility} (ZTF)\footnote{https://www.ztf.caltech.edu/},
the {\it Panoramic Survey Telescope \& Rapid Response System} (Pan-STARRS)\footnote{https://panstarrs.stsci.edu/}, 
and the {\it VISTA Variables in the Via Lactea} (VVV)\footnote{https://vvvsurvey.org/}. 
Also, new eruptive variable candidates are identified almost every month by the 
{\it Gaia Photometric Science Alerts Project}\footnote{http://gsaweb.ast.cam.ac.uk/alerts} \citep{Hod21}. 

The new discoveries have significantly increased the number of potential FUor/EXor variables: only considering 
the Gaia alerts classified as young stellar sources one finds more than 200 cases and a fraction of these 
could well represent good FUor/EXor candidates. 
Although significant, these numbers are however too few to consider eruptive mass accretion as a common 
way through which stars accrete mass. Indeed, non-steady mass accretion has been proposed  \citep{Ken90, Eva09}
as a way to solve the well-known {\it luminosity problem}: the typical observed luminosity of 
the YSOs is at least one order of magnitude lower than expected
if accretion proceeds at rates predicted by the standard steady model \citep{Shu77}.\\
Such a fundamental question will be hopefully answered in the next years thanks to the {\it Legacy 
Survey of Space and Time} (LSST \footnote{https://www.lsst.org/}), which will be performed with the
Vera Rubin Observatory (VRO). This survey will monitor the southern sky in six bands 
(u, g, r, i, z, y) at a 5$\sigma$, single exposure, sensitivity varying between 23.3 and 25.0 
magnitudes\footnote{https://www.lsst.org/scientists/keynumbers}.

This, coupled with an optimal monitoring cadence of a few days,  will allow us to perform for the first 
time statistical studies of eruptive variables based on large samples.

Given the huge number of detected variables expected to be alerted each night, it is mandatory to have 
in place an algorithm capable of both identifying the candidate eruptive variables and distinguishing 
them from other phenomena of stellar variability (e.g. extinction-driven, long-period Mira-type, variability induced by 
magnetic activity, hot/cold spots, or flares). In this paper we describe the algorithm we have implemented 
to identify brightening episodes in generic light curves, 
which has been optimised and tested with a sample of known EXor-type objects, whose typical variability 
timescales are compatible with the measurement cadence and mission lifetime of the LSST.


Our approach relies on the data collected by the public databases cited above 
\citep{Sha14, Jay18, Jay19a,Jay19b, Gai18, Law09, Cha16}, which allow unbiased searches for EXor 
candidates at complementary levels of sensitivity in the optical band, where the intrinsic 
fluctuation of the EXor brightness has the maximum amplitude \citep{Lor07}.

However, even if the light curves considered in the present work are typically sampled 
in days or months, it is important to note that the algorithm presented here is independent 
of the involved time scales because it simply takes into account the ordered sequence of photometric points. 
While on one side this means that the detection of a given feature in the light curve is separated 
from its physical interpretation in terms of the involved time scale, on the other allows a more 
extensive use of the procedure to other kind of light curves. 

In the following sections we will present some typical light curves (Section \ref{sec:lightcurves}) and 
describe the algorithm devised to capture their most relevant features (Section~\ref{sec:method}). 
The implemented procedure is detailed in Section~\ref{subs:pipeline} and the results of its application 
to typical well sampled cases of known or candidate EXors are presented in Section \ref{sec:application}.
Our final remarks are given in Section \ref{sec:conclusions}. \\


\section{Light Curves}   \label{sec:lightcurves}
In the previous section, we have provided a brief description of the main photometric and spectroscopic features 
that characterize the two classes of young eruptive variables, namely the FUors and EXors. This classification 
dates back to some decades ago, when very few objects had been discovered. Although this picture is in general 
still valid, the increasing number of discoveries that occurred in the subsequent years has revealed that 
indeed eruptive variables may exhibit light curves very different (e.g. in duration, cadence, amplitude, 
rising time/decline) from each other. These different properties cannot easily be ascribed to one of 
the two classes. Examples are V1647 Ori \citep{Rei04} and V2493 Cyg \citep{Sem10}, 
whose photometric behavior in terms of outburst amplitude and duration is intermediate between those typical 
of EXors and FUors \citep[e.g.][]{Aud14}. More generally, the photometric surveys cited in the previous section, 
together with long-term monitoring of specific sources \citep[e.g.][]{Hil13, Gia17} and 
archival searches of observations of the last century \citep{Jur17, Jur18}, have revealed 
not only that the light curves of eruptive variables can significantly vary from object to object, but 
also that the same source can be subject to very different outburst episodes for the duration, intensity, and 
speed of brightness increase \citep{Gia20}. \\

It is also worth noting that, in addition to eruptive variables, another kind of young variables exists. 
In these sources, identified as UXor variables, the  brightness fluctuations are (although not only) 
caused by periodic decreases of luminosity due to the transit of an obstacle along the line of sight (e.g. planets, 
dust clumps, irregularities in the circumstellar disks), which causes an enhancement of the circumstellar extinction 
and brightness variations of similar amplitude and cadence as those of accretion-driven variables. However, UXors 
often show a periodicity in the light curve, while accretion bursts are episodic events with an apparently stochastic recurrence.

\begin{figure}
\center{\includegraphics[width=1.\textwidth,height=0.65\textheight]{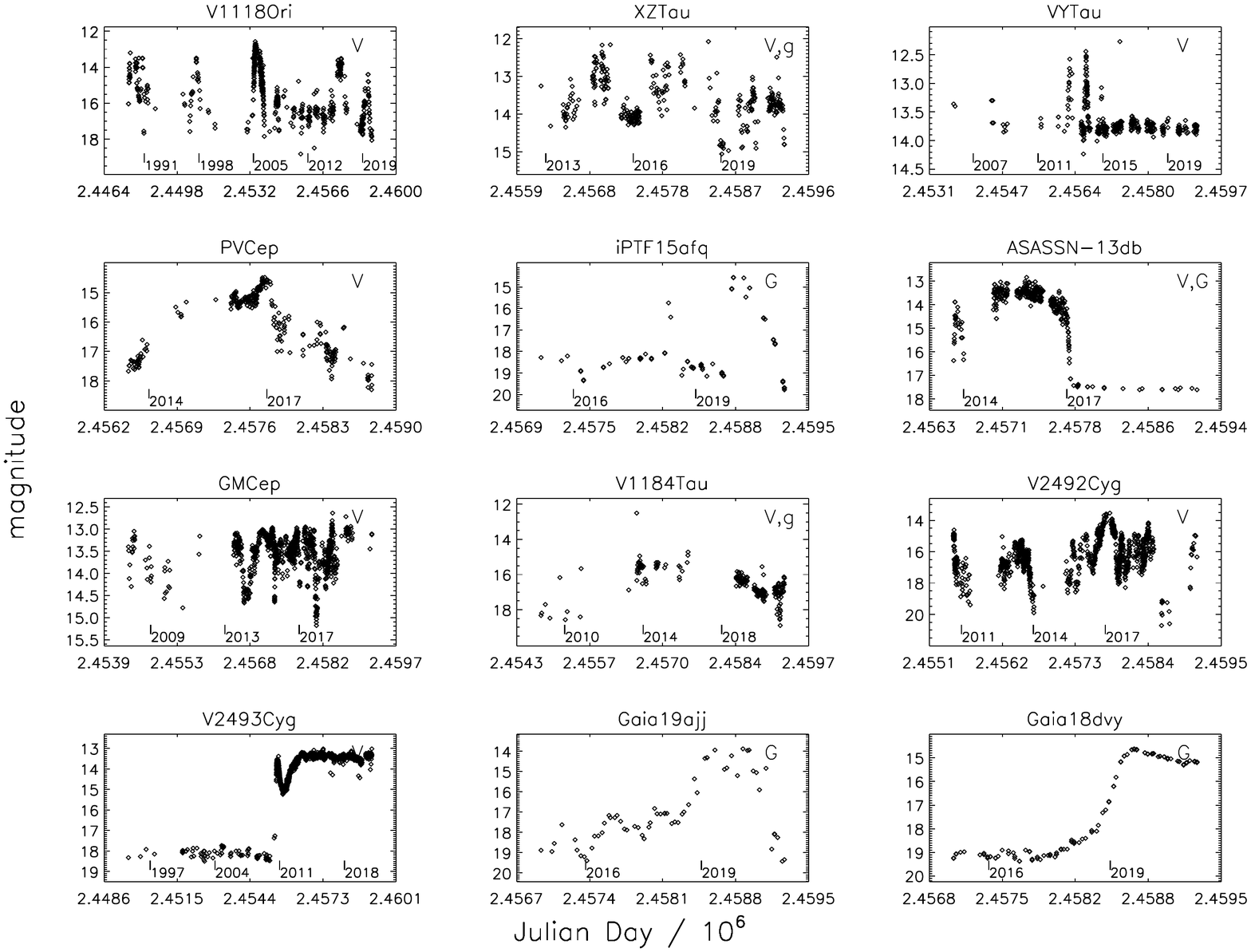}} 
\caption{Examples of light curves illustrating the wide variety of photometric behaviors 
         exhibited by eruptive young stars. The additional tick marks at the bottom 
         of each panel correspond to the beginning of the indicated years in the light curves. 
         For each object, the photometric band of the observations is shown in the upper right corner. 
         The light curve of V1118 Ori in the V band has been collected from the literature, while 
         for XZ Tau, ASASSN-13db, and V1184 Tau the photometry has been assembled by using two 
         input catalogs (see text). \label{fig1_lc_all}} 
\end{figure}

In Fig.~\ref{fig1_lc_all} we show some light curves to exemplify the different behaviors described above. All the data 
have been retrived from the ASAS-SN ($V$-band), the ZTF ($g$-band), or the Gaia ($G$-band\footnote{The Gaia Alerts G magnitudes are derived from a preliminary
 calibration of the Gaia unfiltered (white light) photometry. The Gaia ``G'' band is defined by the instrument response only
 and covers a broad wavelength range,  from the near ultraviolet ($\sim$330 nm) to the near infrared($\sim$1050 nm).}) databases, with the exception 
of \astrobj{V1118 Ori} whose data ($V$-band) are also taken from the literature \citep{Gia20}.  
The time coverage of the photometry depends on the source, spanning in the range from $\sim$5 years (e.g. \astrobj{PV Cep}) to $\sim$30 years 
(\astrobj{V1118 Ori}), and is shown in Table~\ref{table1} along with other information on these objects. \\

Since the light curves are shown here only for illustrative  purposes, we did not apply any photometric correction to 
convert the data points to a unique band. This procedure, despite rough, allows us to obtain a broader view of the 
photometric behavior of these objects. In particular this is the case for XZ Tau and V1184 Tau, where 
the available ASASSN observations obtained in V band (central wavelength $\sim$5500 \AA) have been complemented with 
Gaia G ($\sim$5800~\AA) data, and for ASASSN-13db whose light curve has been supplemented with ZTF g ($\sim$4800 \AA) photometry 
approximately covering the last two years. 


In the first three panels, we show the light curves of three classical EXors: V1118 Ori, XZ Tau, and VY Tau. The first two 
objects show a similar, although not identical, behavior, with  stochastic brightness increases of about 3 magnitudes, 
typically lasting several months that are interspersed by periods of quiescence during which a certain level of activity 
is however still recognizable.

Conversely, VY Tau presents a well-defined level of quiescence  around $V$\,=\,13.7 mag 
interrupted by sudden and very short episodes of significant brightness increase. The fourth panel shows how the 
light curve of PV Cep passed from the quiescence to the outburst phase very slowly (about 3 magnitudes in 3 yr)
then started declining again after 2017. In the fifth and sixth panels are depicted the light curves of two sources, 
iPTF15afq and ASASSN-13db, that show rapid brightness increase ($\sim$ 3 mag in a few months) followed by a 
comparatively rapid decrease. In particular ASASSN-13db is reminiscent of both the behaviors observed in V1118 Ori 
and in PV Cep: it has undergone a first outburst of short duration and high speed in both the rising and declining phases, 
followed by a second outburst of higher intensity and much longer duration. 

In the next three panels of Fig.~\ref{fig1_lc_all} the light curves of some objects usually classified 
as UXors are presented: 
GM Cep, V1184 Tau, and V2492 Cyg. Out of these, GM Cep shows an 
almost stable high-level state with roughly periodical dimmings, while V1184 Tau presents a slowly increasing 
and then decreasing light curve with superimposed lower amplitude variations that could be ascribed to 
accretion events \citep{Gia16}.  V2492 Cyg  also shows periodical, semi-regular variations that are likely due 
to either extinction or accretion phenomena \citep{Hil13}. \\

Finally, in the last three panels, there is the light curve of V2493 Cyg, a source whose behavior is intermediate 
between an EXor and a FUor, followed by two examples of YSO variables recently discovered by Gaia. The first (Gaia19ajj) 
looks like PVCep or V2492Cyg, increasing of about 5 mag in five years between 2015 and 2019 and rapidly decreasing 
to the presumably quiescent state in one year, while the second (Gaia18dvy) has undergone a slow, monotonical increase 
between 2016 and 2019, passing from G $\sim$19 to G $\sim$15. Since 2019, this source is fading again, but at an 
even lower rate.

\begin{table}[h]
\footnotesize
  \caption{Characteristics of the sources$^a$ in Fig.~\ref{fig1_lc_all}. \label{table1}}
\begin{threeparttable}[h]
  \centering
  \begin{tabular}{lccccccc}
  \hline\hline 
     Name    &  VarType   &  Ref$^b$   & Simbad class    & photometry source     &  time coverage & bursts detected \\
             &            &            &                 &                       &   (yr)         &                 \\
\hline
 XZ Tau      &  EXor ?    &  1         & Orion &   ASASSN+ZTF    & 8.3  & 2  \\
 VY Tau      &  EXor      &  1         & Orion &   ASASSN        & 15.1 & 0$^c$ \\
 ASASSN-13db &  EXor      &  2         & transient & ASASSN+Gaia & 7.0  & 2  \\
 V1118 Ori   &  EXor      &  1         & Orion &  \citet{Gia20}  & 31   & 6  \\
 V1184 Tau   &  UXor/EXor &  1         & Orion &   ASASSN+ZTF    & 12.2 & 1  \\
 iPTF15afq   &  EXor ?    &  3         & YSO   &   Gaia          & 5.9  & 2  \\ 
 Gaia19ajj   &  EXor ?    &  4         & YSO   &   Gaia          & 6.3  & 1 \\
 Gaia18dvy   &  EXor      &  5         & Orion &   Gaia          & 6.1  & 1 \\
 PV Cep      &  EXor ?    &  1         & Orion &   ASASSN        & 6.3  & 1  \\ 
 V2492 Cyg   &  EXor ?    &  1         & Orion &   ASASSN        & 10.4 & 3  \\
 V2493 Cyg   &  EXor/FUor &  6         & Orion &   ASASSN        & 26.5 & 1 \\  
 GM Cep      &  UXor      &  7         & Orion &   ASASSN        & 13.1 & 1$^d$  \\     
\hline
  \end{tabular}
\begin{tablenotes}
\footnotesize
\item[(a)] ordered in increasing Right Ascension.
\item[(b)] (1) \citet{Aud14}; (2) \citet{Hol14}; (3) \citet{Mil15}; (4) \citet{Hil19}; (5) \citet{Sze20}; (6) \citet{Kos11}; (7) \citet{Sem10}
\item[(c)] while bumps are recognizable, they are not detected as bursts because $\Delta\textrm{m} < 2$ mag.  
\item[(d)] probably fictitious because the source resembles the UXor type, showing fadings more than bursts.
\end{tablenotes}
\end{threeparttable}
\end{table}

Given the variety of light curves exhibited by young eruptive variables, we have decided to devise a recognition 
algorithm based on the observational data of V1118 Ori, which is one of the very few objects for which the light curve has 
been frequently sampled and monitored for tens of years. Therefore, our procedure follows two main steps: 1) implementation 
of an algorithm able to recognize all the bursts observed in V1118 Ori; 2) validation of this algorithm to recover 
the accretion outbursts observed in the light curves of other objects, also to test its capabilities and limitations.

\section{Adopted Method} \label{sec:method} 
We are interested in the detection of EXor-like candidate events because the 
light curves of this kind of sources are known for only a handful of well sampled cases. 
In this context, statistical approaches, including advanced light curve analysis for 
transients, cannot be reasonably trained and validated before to be used. 

We have developed an ``ad hoc'' procedure to detect potentially 
interesting cases, namely sources exhibiting brightenings of the kind expected for 
candidate EXors. Indeed, here we focus our attention on the detection of bursts in a light curve, 
leaving the classification of the sources to further considerations (as burst duration, recurrence, 
rising or declining speed, etc.) that are out of the scope of this work. 

Here the approach is in fact both simple and empirical, meaning that the sequence 
of steps implemented in our procedure closely resembles what a trained eye 
would suggest in classifying a given sequence of photometric points that progressively 
build up the light curve. 

The method is based on the recognition of a reference observation, hereinafter 
called ``reference point'' which can play the role of the {\it bona fide} 
quiescence level of the source. An important characteristic of this point is that it remains a 
fixed point until a new observation is identified which is deemed to be more 
representative of a quiescent state. In such a case the reference point 
is updated by correspondingly assigning the reference status to the new recognized position 
in the light curve.

In practice, the first observation of a light curve is adopted as the initial reference point. The analysis 
then proceeds point by point, until the subsequent photometric behavior of the light curve 
suggests that the reached brightness can represent a better estimate of the quiescence 
level. In this latter case the reference status is assigned to the newly deemed quiescent point. 
This algorithm essentially compares the magnitude of each new point in the light curve with 
the magnitude at quiescence and, if the difference exceeds a given threshold, the source 
is considered to be in a higher brightness state. 

These basic considerations have been implemented in an IDL procedure, hereinafter 
called pipeline, aimed to take into account several possible behaviors exhibited 
by a set of points empirically chosen in the final tail of the light curve. 

The pipeline goes sequentially through the photometric series, evaluating each point 
with respect to the characteristics of the previous light curve for the purpose 
of detecting specific behaviors. To this aim a sequence of steps have been 
devised in such a way that, when a given condition is fulfilled, an appropriate flag is assigned 
to the point and the analysis moves on to consider the next photometric point. \\

\subsection{Pipeline description} \label{subs:pipeline}
To assist the description of the steps involved in the pipeline, in Fig.~\ref{fig2_cdl_artificial} we show an artificial 
light curve simulating typical photometric behaviors we expect to find in eruptive variable sources. 
Even if the sampling interval adopted in this simulated curve is of the order of the days, the evaluation of 
the state of the source will be based merely on the observational sequence of points, leaving any 
consideration on the timescales involved to a subsequent physical interpretation of the brightness 
variations. Because of this we shall focus our attention on the simple detection of some interesting 
characteristics of the light curves: 
\begin{itemize}
\item[a)] a reference point representing a plausible baseline (quiescence level); 
\item[b)] short-term magnitude changes of a duration less than the sampling interval (e.g. a spike); 
\item[c)] long-term magnitude changes extending over multiple observations (bursts or fadings);   
\item[d)] drifts in the brightness of the baseline. 
\end{itemize}

A web interface for our pipeline is accessible at \href{https://exorlc.le.infn.it}{https://exorlc.le.infn.it}.

\subsubsection{Pipeline details}
Here the implementation details are described and since this section is quite technical 
it can be skipped without compromising the comprehension of the subsequent sections. 

The light curve analysis is carried out in a loop that sequentially examines all the observations through a series of steps. 
In our approach an important role is assigned to what we call ``reference point'' that  
represents our best guess, given the available data in the evolving light curve, of the source quiescent state. 
Consequently, an essential part of the pipeline is oriented to the identification of this point 
that stands for the reference magnitude against which the algorithm evaluates the brightness changes.
However, because even in quiescence our sources are typically characterized by some more or less important 
brightness fluctuations, we introduce a tolerance parameter quantifying the amplitude of such 
intrinsic variations. This can then be used to adapt the detection algorithm to the kind of light curve 
analyzed.

In the following we describe the considerations we adopted to empirically classify a given point $i$ 
in one of the possible states, namely: reference (estimated quiescence), high, drop, spike, generic. 
The implementation of the pipeline begins by defining the starting observation in the light curve as 
the first reference point, and then skipping a few subsequent points whose status is difficult to 
estimate on the basis of a still too short light curve. The procedure then follows a sequence of 
steps aimed to evaluate at each point the corresponding status, according to these considerations:

\begin{figure} 
\center{\includegraphics[width=0.8\textwidth,height=0.3\textheight]{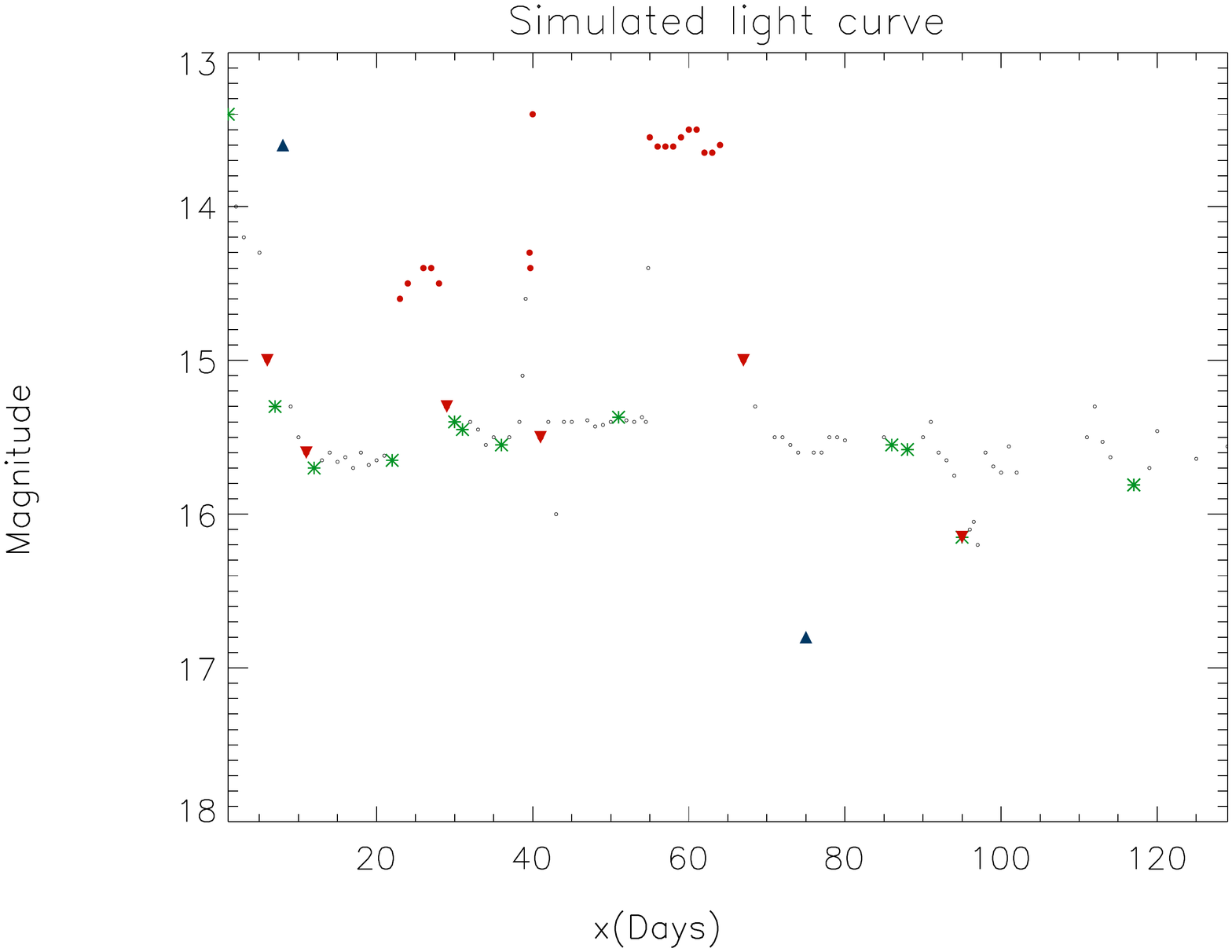}} 
\caption{Simulated light curve adopted for the detection of brightenings larger than 1 mag. 
The points of the estimated quiescence level (reference points) are shown with a green asterisk, 
while those corresponding to a high state phase are shown as red filled circles. A blue triangle 
indicates a spike, namely a point considerably deviating from the immediately adjacent points, 
while red upside-down triangles mark remarkable drops in brightness. \label{fig2_cdl_artificial}}
\end{figure}

\begin{itemize}
   \item[1-] {\it Reference point sliding toward the local minimum of brightness}. 
             If the current point $i$ is immediately preceded by an already defined reference point in $i-1$ 
             and if the source in $i$ is fainter than in $i-1$, then it is chosen as the new reference. 
             In Fig.~\ref{fig2_cdl_artificial} this case produces the two green asterisks at x=32 
             and x=88 which mark points that, satisfying this criterium, relocate the previous reference point
             and ensure that the estimated quiescent level actually corresponds to a local minimum.

   \item[2-] {\it Reference point before a relevant brightening.}
             In this case the sequence of the four last points $i-3, i-2, i-1, i$ is considered. 
             If they show a monotonically increasing brightness and the accumulated magnitude difference 
             is larger than a given threshold, 
                  $ m_{i-3} - m_i > \Delta \textrm{m}_{\textrm{th}} $,
             then the reference is updated if in the point $i-3$ the source is fainter (within a given tolerance) 
             than in the preceding reference point. 
             Such a case is illustrated in Fig.~\ref{fig2_cdl_artificial} by the green asterisk just before 
             the brightening at x=22. 

   \item[3-] {\it A plateau in the light curve suggests a quiescent state and then a new reference}. 
             When the magnitude of the point $i$ is consistent with the mean magnitude 
             of a reasonable number of immediately preceding points (here we adopt 7 points) 
             whose standard deviation is less than 0.1 mag, then the point is considered on a plateau 
             of the light curve. In this case, if the magnitude of the preceding reference point is larger 
             than (within the tolerance) the plateau mean magnitude, the point $i$ is considered as the new 
             reference point. This is the case illustrated 
             in Fig~\ref{fig2_cdl_artificial} by the green asterisk at x$\sim$ 50.

   \item[4-] {\it A minimum in a descending branch could represent a reference point }. 
             If the sequence of points $i-3, i-2, i-1$ exhibits a continuously decreasing brightness, while  
             the point $i$ shows instead an increase, then a reference status is assigned to the point $i-1$ 
             if  $m_{i-1} > m_{\textrm{ref}}$, even considering the tolerance, as exemplified in 
             Fig.~\ref{fig2_cdl_artificial} at position x=95;

   \item[5-] {\it Detection of spikes}. A ``spike'' is flagged when in a sequence of five points 
          the magnitude of the midpoint differs more than a predefined threshold from the 
          linear fit to the 4 adjacent points $i-4, i-3, i-1, i$. This definition captures single points 
          significantly deviating from the neighborhood and includes both upward and downward brightness changes. 
          Should the midpoint $i-2$ have been previously assigned to another status (e.g. reference or high), 
          then this assignment is removed and the point is flagged as a spike. In the case the previous 
          assignment was a reference, the immediately preceding reference point is recovered for the subsequent analysis. 
          Two spikes are visible in the simulated light curve at the positions x=8 and 75 

   \item[6-] {\it Sudden brightness drop and update of the reference point}. 
          If the magnitude in the last two points, $i-1$ and $i$, is larger than in the reference and 
          the difference $m_i-m_{\textrm{ref}}$ is larger than a predefined threshold, 
          the point $i$ is highlighted as a ``drop'' due to the sensible fading of the source. 

          These cases are shown as upside-down red triangles (x=6, 12, 95).  
          Furthermore the point is also taken as a reference point if one of these conditions is satisfied:

      \begin{itemize}
         \item[i)]    the magnitude in $i-3,i-2,i-1,i$ grows monotonically and the magnitude difference 
                      $m_i-m_{i-3}$ is larger than the tolerance (e.g. at x=95); 

         \item[ii)]   the last five points of the light curve are cradle-shaped, free from spikes, 
                      and the magnitude difference $m_{i-2}-m_i$ is larger than the tolerance. 
                      If this condition is not satisfied, a further point $i-5$ is considered to 
                      verify if its magnitude is coherent with the cradle-shaped trend and the difference 
                      $m_{i-2}-m_{i-5}$ is larger than the tolerance. 
      \end{itemize}
          In the same line a point is similarly highlighted as a drop when the sequence of points $i-3, i-2, i-1, i$ shows a  
          monotonically descending brightness, so that the accumulated difference in magnitude is larger 
          than a threshold (e.g. at x=29, 41, 67). Even in cases of non-monotonic fading, the point is anyway 
          considered as a drop if the magnitude difference between the points $i$ and $i-4$ is larger than the threshold.

   \item[7-] {\it Assign the ``high'' status to the point}. 
             In the opposite case in which the magnitude at both $i-1$ and $i$ points is significantly lower than in the 
             last reference point and the difference $m_{\textrm{ref}}-m_i$ is larger than a predefined threshold, 
             the point $i$ is flagged to represent a high state and is shown in Fig.~\ref{fig2_cdl_artificial} as a red point.
             If instead the points $i-1$ and $i$ fall below the threshold, the behavior of the light curve is 
             further analyzed and a reference point is assigned if one of these conditions is verified: 
        \begin{itemize}
             \item[-] the source in the last points is continuously and significantly fading, with 
                      the last point falling, within the tolerance, near the magnitude of the last reference point;
             \item[-] the sequence of the last five points is cradle-shaped with the lower brightness 
                      midpoint beyond the tolerance from the magnitude of the last reference point. 
                      An example of this being the reference point marked in x=117.
        \end{itemize} 

\end{itemize}

%
%

\subsection{Tests on a known source} 
\begin{figure}[h]
\centerline{   \includegraphics[width=0.8\textwidth,height=0.5\textheight]{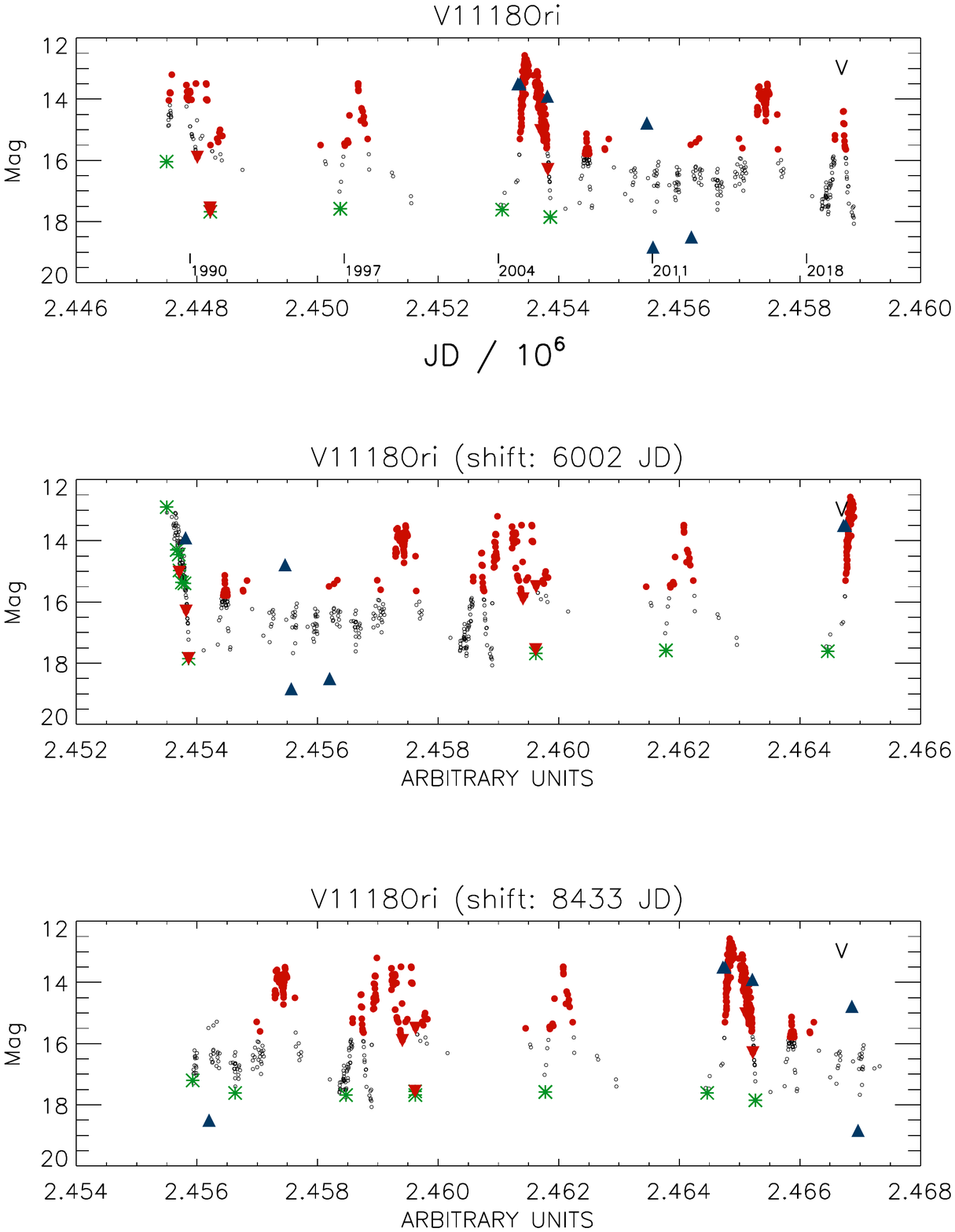}}
   \caption{The V1118~Ori light curve adopted for testing the pipeline. Symbols and colors are as 
            in Fig.~\ref{fig2_cdl_artificial}. Upper panel: the result of the analysis 
            of the V1118 Ori light curve with the time axis in the same units as in Fig.~\ref{fig1_lc_all}.
            Second and third panel: light curves modified for testing 
            purposes, obtained by assuming a different starting point. The x-axis 
            values correspond to the observations only in the upper panel, the other two panels 
            being artificially obtained by x-shifting the first point of the original curve (see text). \label{fig3_v1118ori_shifted}} 
\end{figure}

To illustrate the results obtained by experimenting the pipeline on an actual light curve, 
we used the data collected for V1118~Ori, probably the best observed EXor source, whose 
light curve is shown in the upper panel of Fig.~\ref{fig3_v1118ori_shifted}, encompassing 
about 30 years from $t_{\textrm{ini}} \simeq 1989$ to $t_{\textrm{fin}} \simeq 2020$. 
In this figure 
red filled circles are used to emphasize the high state phase of the  source, corresponding to a 
brightening of $\Delta{\textrm{m}} > 2$ mag with respect to the estimated quiescent state 
(see Section~\ref{subs:pipeline}), the latter represented by the preceding green asterisk. 
A blue triangle is used to mark a spike that here corresponds to a jump of $\Delta m= \pm$ 1 mag.

Since the ligh curve analysis begins by considering the initial observation as the first reference 
point, it is necessary to ensure that the results of the pipeline are reasonably independent of the 
particular state of the source at the beginning of the observations. 

To this aim we modified the light curve by moving the initial point at $t_{\textrm{ini}}+\Delta t$, 
artificially appending all the skipped data points after the last observation at $t_{\textrm{fin}}$, 
namely at the end of the original curve. In this way we obtained a number of artificial light curves 
that, once submitted to the pipeline, always show the same points recognized as high state, except in the 
very initial part of the curve where the results depend on the stochastic position chosen for the 
starting point that, in turn, defines the first reference point. 

This is illustrated by the remaining two other panels in Fig.~\ref{fig3_v1118ori_shifted} that 
essentially show a replica of the same light curve, but with the initial point assumed in 
two different dates corresponding to an active and a quiescent state, respectively. 

In the middle panel a shift of $\Delta$t=6002 JD brings the peak of the 2005/06 bump in the 
first position and consequently the bump itself remains unrecognized due to the lack of a previous 
reliable reference point actually associated to a quiescent state. This is progressively better devised in the 
subsequent descending phase of the brightness, until the bottom of the curve is reached where a 
new reference point is assigned, representing more faithfully the quiescence. Clearly, from this position onward, 
the analysis of the light curve closely follows the results obtained for the original curve in the top panel. 

The last panel at the bottom shows the same light curve, but shifted by $\Delta$t=8433 JD so that 
the starting point of the observations falls now in a low activity region of the original curve. 
Because of this, the assumed first observation is already near a quiescent phase of the source so 
that all the subsequent points are classified similarly to the original curve. 

Whatever the shift we adopt, after a transitory initial phase needed to estimate a representative quiescent 
level, the analysis always detects the same high states making us confident on the capability 
of the pipeline to reasonably find the high state episodes of the source. This also means that 
an alert can be appropriately issued, should the increase in brightness with respect to the 
reference point exceed a given threshold. Indeed, repeating this kind of test with a random initial point, 
the results are invariably very similar once the reference point has reached a reliable quiescent state, 
with the result that all the bursts reported in the literature for this source are always detected. 
The exception is clearly the case in which the beginning of the light curve falls on a burst that, 
in this way, is missed as illustrated in the 2nd panel of Fig.~\ref{fig3_v1118ori_shifted}.

\section{Application to case examples} \label{sec:application}

\renewcommand{\thefigure}{4a} 
\begin{figure}
\includegraphics[width=0.9\textwidth,height=0.6\textheight]{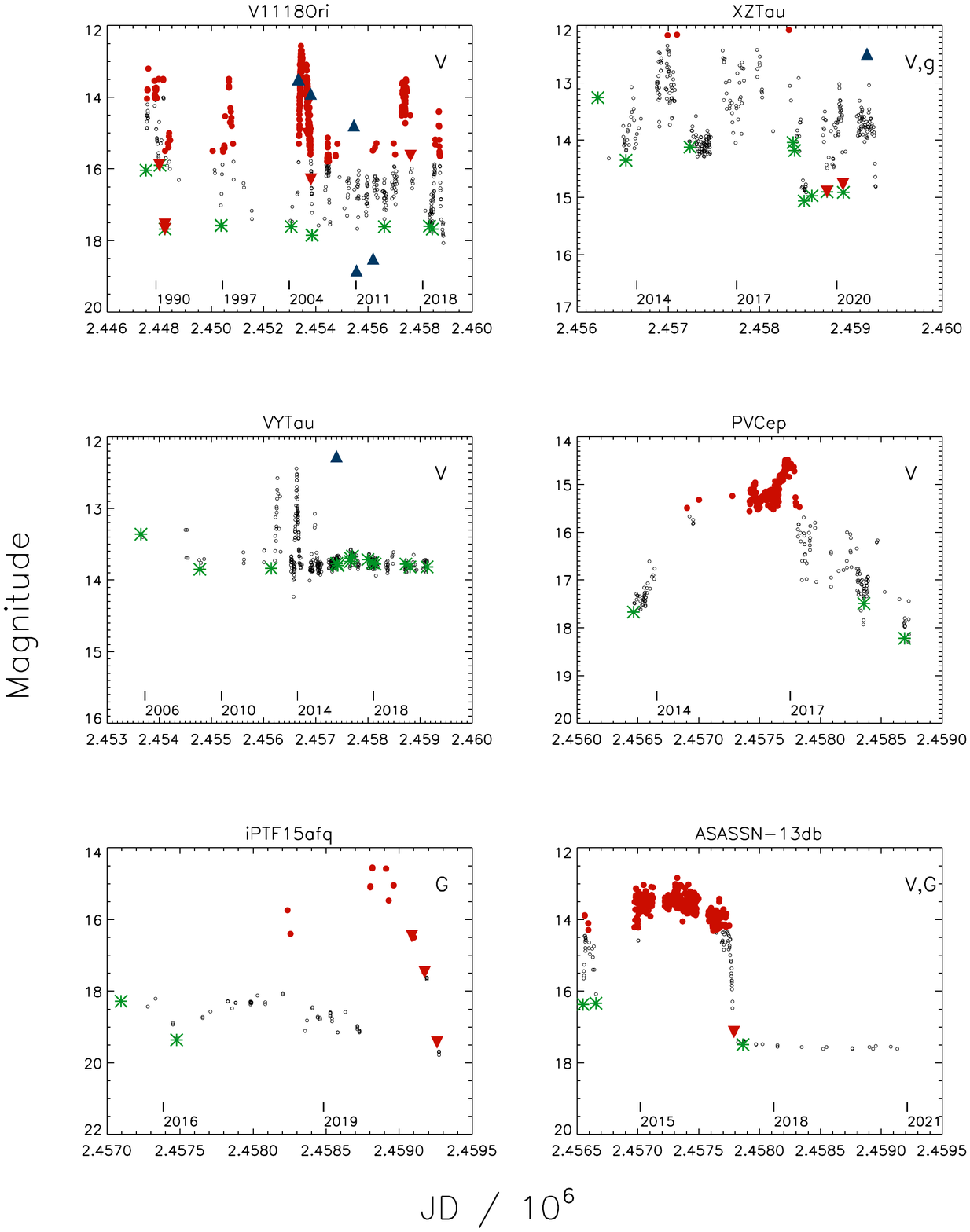}
\caption{The light curves of the first six sources in Fig.~\ref{fig1_lc_all}, 
   with superimposed the reference (green), spike (blue), and high state (red) points 
   established by using the pipeline described in Section~\ref{subs:pipeline}. In each panel, the spectral 
   band of the observations is shown with a letter in the upper right corner, two letters being 
   used to indicate the use of two different bands (see text). The threshold in magnitude and the tolerance 
   have been set to 2 and 0.2 mag, respectively. 
   \label{fig4_lc_detections}}
\end{figure}

With the aim of testing the procedure against other photometric behaviors, we submitted the light curves 
in Fig.~\ref{fig1_lc_all} to our pipeline, obtaining the results shown in Fig.~\ref{fig4_lc_detections} 
and \ref{fig5_lc_detections}. 
In these figures the threshold in the magnitude difference required for a point $i$ to be considered 
as ``high state'' has been set to $\Delta m = m_\textrm{ref} - m_{i}~>~2$~mag, quite typical of EXor sources. 
The tolerance has been taken at 0.2 mag to account for the small brightness fluctuations considered as intrinsic 
to the quiescence state. For simplicity, here we adopt the same value for all the analyzed sources even if, in principle, 
this parameter could be tuned on specific values tailored on the small fluctuations actually 
present in the light curve of each source (see Section~\ref{subs:pipeline}). 
Finally, the threshold for a spike has been set to 1~mag.

In Fig.~\ref{fig4_lc_detections}, the upper left panel is taken from Fig.~\ref{fig3_v1118ori_shifted} 
and is reported here for ease of comparison with the other sources. Note that at the beginning of the year 1990 
there is a notable fading in brightness that is marked as both reference point (green asterisk) and drop 
(red upside-down triangle, overplotted), 
the latter introduced to highlight rapid brightness decreases (see Section~\ref{subs:pipeline}, point 6). 
Similar cases are also found in iPTF15afq in the descending branch at the end of the light curve and in ASASSN-13db 
in the rapid fading at the end of 2016. In Fig.~\ref{fig5_lc_detections} the subsequent  two other cases are also 
evidenced: V2492 Cyg whose brightness decreased $\sim$5 mag in 2018, and Gaia19ajj that experienced a flux 
drop of $\sim$3 mag toward the end of 2020. \\

In the case of XZ Tau the fluctuations of the light curve are always within the adopted 2 mag threshold, with the 
exception of three isolated points, two of which located around the small brightening peak seen between the end 
of 2014 and the beginning of 2015, and the other high point being more isolated. Despite the latter could resemble a spike, this 
is not recognized as such by the pipeline because of its position is compatible with the ordered sequence of the 
immediately preceding and following points. 
A spike is instead highlighted (see Section~\ref{subs:pipeline}, point 5) at the end of 2020, shown as a blue triangle.\\

The next source, VY Tau, shows an approximately constant baseline at $\sim$13.7~mag, interrupted by three small and sharp 
peaks suggesting that this object suffers rapid and isolated jumps in brightness of the order of 1~mag. Their amplitude is 
however too small with respect to the adopted threshold, so that they are not highlighted by the pipeline. There 
is also a point marked as a spike that, given its brightness and time of occurrence, could be interpreted as a fourth 
rapid brightening belonging to the same series of episodes. However, due to its isolation with respect to the adjacent points, 
it has been formally marked as a spike. \\

The light curve of PV Cep shows instead an almost monotonic rise (years 2013--2015) until the high state phase 
is reached lasting $\sim$2 years (2015--2016). The subsequent decline appears much more bumpy and without 
peculiar points as spikes or drops.\\

As in the case of VY Tau, even the light curve of iPTF15afq shows an almost stable baseline (mag $\sim$18) in 
the years 2015--2019 with only two adjacent points highlighted by the pipeline. At the end of 2019 the brightening was 
$\sim$4~mag and lasted a few months, while in the year 2020 a larger and rapid fading ($\sim$5~mag) has been observed 
lasting approximately one year and followed by a comparatively slower decline. A sequence of red upside-down 
triangles marks a series of drop in brightness that has now reached $\sim$20~mag. \\

In the last panel the ASASSN-13db light curve begins at the end of 2013 with a 100 days bump, with the brightness rapidly 
rising in 20 days at a rate of $\sim$0.1 mag/day and then falling at $\sim$0.025 mag/day to the previous level of $\sim$16 mag. 
Then, after a one year gap in the observations, the source is found again in a high state at $\sim$14 mag, remaining bright for 
more than two years until January 2017 when brightness falls at a rate of 0.1 mag/day reaching a plateau 
at $\sim$18 mag. At the end of this rapidly descending phase a red triangle marks the point as 
a ``drop'' indicating a sensible decrease in brightness with respect to the preceding observations 
(see Section~\ref{subs:pipeline}, point 6).\\

\renewcommand{\thefigure}{4b}
\begin{figure} 
\includegraphics[width=0.9\textwidth,height=0.6\textheight]{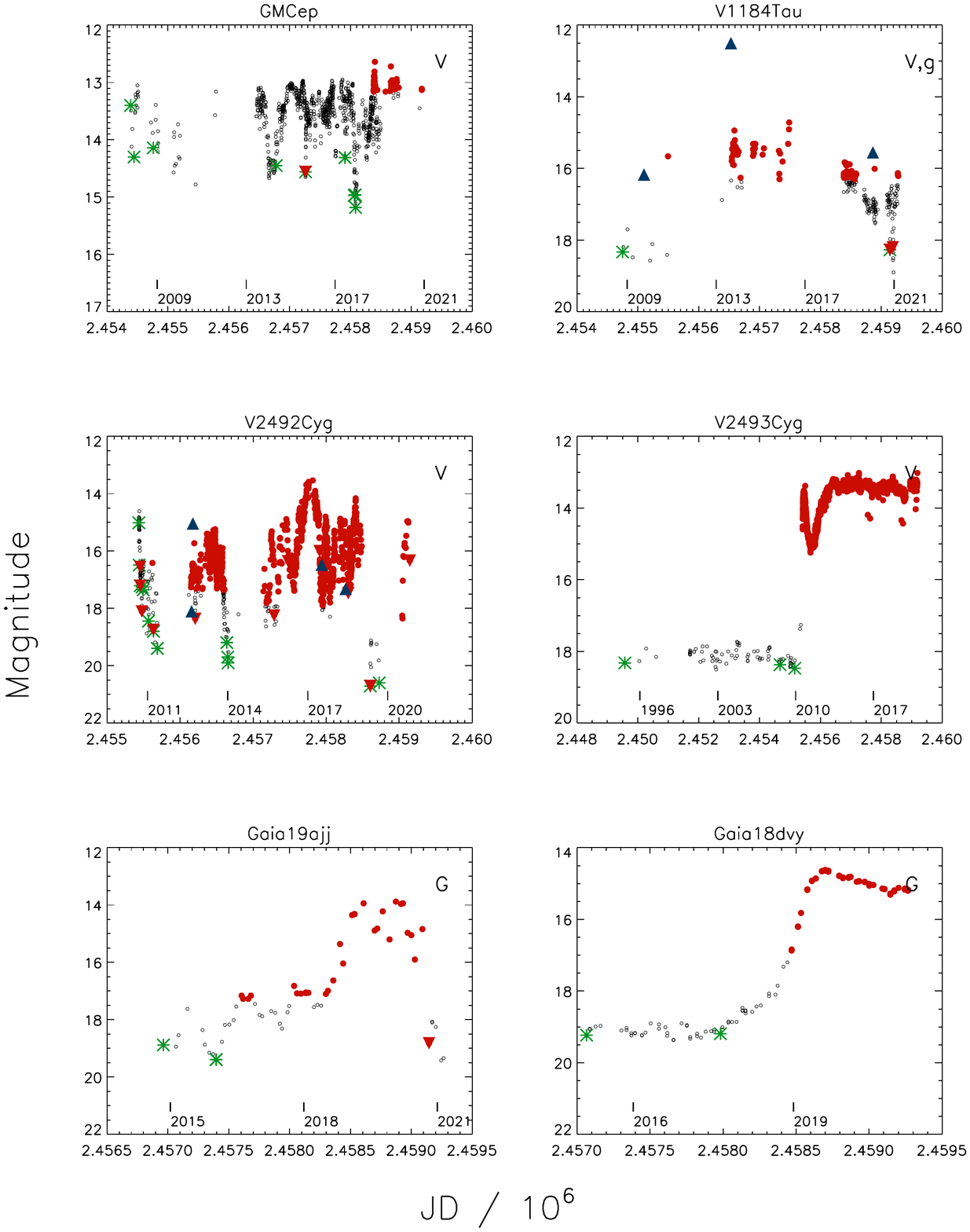}
\caption{As in Fig.~\ref{fig4_lc_detections}, but for the last six sources of Fig.~\ref{fig1_lc_all}. \label{fig5_lc_detections}}
\end{figure}

The results obtained in the remaining six example cases are shown in Fig.~\ref{fig5_lc_detections}. 
In the upper left panel the light curve of GM Cep is presented with some 
points highlighted toward the end of the curve. Their magnitude difference from the preceding reference point  
is actually larger than the threshold but, should the reference point represents more an extinction variation 
than a quiescent level, the high state points could be more apparent than due to a genuine 
activity of the source. \\

The next case, V1184 Tau, appears dominated by a global trend encompassing more than $\sim$10 years. 
In the first part of the light curve the brightness increases slowly for more than 2 mag, 
albeit with large fluctuations, surpassing the adopted threshold ($\Delta m = 2$ mag) in 2013 and then, 
after $\sim$2 years, begins to decline reaching $\sim$18 mag. In this evolution, spikes are detected in 2009, 2013,
and 2019, and a drop, more recently, at the end of 2020.\\

The two subsequent light curves show very different behaviors. V2492 Cyg appears irregularly 
variable and the pipeline, besides highlighting the high states, finds several spikes and drops in brightness. 
On the contrary, V2493 Cyg maintains a quite constant brightness in the years preceding 2010, jumping 
of $\sim$5~mag in six months and then remaining in a high state till now. \\

The last two cases analyzed are Gaia sources, both showing an increase in brightness of $\sim$5~mag in three years. 
However, Gaia19aij has the light growing phase accompanied by large fluctuations of $\sim$1 mag and, after 
reaching a high state, in 2020 rapidly decreases as indicated by the red upside-down triangle. On the contrary, 
Gaia18dvy grows in brightness much more regularly so that no peculiar points are detected, until the magnitude 
difference with the reference point exceeds the adopted threshold and the object is considered in high state. \\

A summary is given in Table~\ref{table1} that, besides showing the time intervals covered by the light 
curves analyzed, also reports the number of detected bursts, namely clusters of connected points recognizable 
as a single brightening episode with amplitude $\Delta m > 2$ mag.\\

   However, we underline that using this algorithm is clearly not enough to pick-up a specific kind of 
   object because our procedure only represents a first step, aimed to select candidate sources to be 
   subsequently scrutinized with additional criteria (e.g., burst time scale, rising and declining slope, 
   repetition) specialized on the observational characteristics of the specific class of objects. 

\section{Conclusions} \label{sec:conclusions}

Nowadays large number of well-sampled light curves are being collected. 
The trend will grow as the next generation of survey telescopes, both ground and 
space-borne, come into operation. 

This new opportunity is ideally suited to answer in an unbiased way 
important questions about the evolution of the pre-main sequence stars, and in particular 
of their disks which are in the planet forming phase. With the aim to have a tool suitable to 
scrutinize large amounts of light curves, we have developed a simple software capable of 
searching for rapid brightening and fading in light curves, 
particularly for the detection of the EXor phenomena to shed light on their 
sporadic accretion processes as the origin of stellar masses. \\

The pipeline has been tested on a few well-sampled light curves 
of recognized or candidate EXors with encouraging results. 
For our tests we adopted a threshold of $\Delta m$=2 mag to detect a burst in 
brightness, a typical value characterizing the PMS stars classified as EXors. 
Finally we emphasize that this work is devoted to the simple detection 
of bursts in light curves, leaving the problem of the source classification 
to subsequent analysis. \\

\section{Acknowledgments}
We acknowledge ESA Gaia, DPAC and the Photometric Science Alerts Team (http://gsaweb.ast.cam.ac.uk/alerts).
We also thank the Computing Service of the INFN-Sezione di Lecce for assisting us in preparing the web interface.




{}

\end{document}